\documentclass[12pt,a4paper]{article}

\usepackage[british]{babel}

\usepackage[a4paper,top=2cm,bottom=2cm,left=2.5cm,right=2.5cm,marginparwidth=1.75cm]{geometry}

\usepackage[style=apa, backend=biber, maxcitenames=2, maxbibnames=999]{biblatex}
\usepackage{amsmath}
\usepackage{graphicx}
\usepackage[colorlinks=true, allcolors=blue]{hyperref}
\usepackage{hyperref}
\usepackage[title]{appendix}
\usepackage{mathrsfs}
\usepackage{amsfonts}
\usepackage{booktabs} 
\usepackage{caption}  
\usepackage{threeparttable} 
\usepackage{algorithm}
\usepackage{algorithmicx}
\usepackage{algpseudocode}
\usepackage{listings}
\usepackage{chngcntr}
\usepackage{booktabs}
\usepackage{lipsum}
\usepackage{subcaption}
\usepackage{authblk}
\usepackage[T1]{fontenc}    
\usepackage[pagewise]{lineno}
\usepackage{csquotes}       
\usepackage{diagbox}
\usepackage{comment}
\usepackage{enumerate}

\usepackage{setspace}
\usepackage{titlesec}
\titleformat{\section} 
  {\normalfont\Large\bfseries}{\thesection.}{1em}{}
  
\usepackage{lineno} 

\leftlinenumbers 

\usepackage{float}   
\usepackage{caption} 


\title{Improving Oil Slick Trajectory Simulations with Bayesian Optimization}

\author[2,1]{Gabriele Accarino} 
\author[1,4]{Marco M. De Carlo} 
\author[1,*]{Igor Atake} 
\author[1]{Donatello Elia} 
\author[1]{Anusha L. Dissanayake} 
\author[6]{Antonio Augusto Sepp Neves} 
\author[6]{Juan Peña Ibañez} 
\author[1,3]{Italo Epicoco} 
\author[1]{Paola Nassisi} 
\author[5]{Sandro Fiore} 
\author[1]{Giovanni Coppini} 

\affil[1]{\small CMCC Foundation - Euro-Mediterranean Center on Climate Change, Italy}
\affil[2]{\small Department of Earth and Environmental Engineering, Columbia University, New York, U.S.A.}
\affil[3]{\small Department Engineering for Innovation, University of Salento, Italy}
\affil[4]{\small National Institute of Oceanography and Applied Geophysics - OGS, Italy}
\affil[5]{\small Department of Information Engineering and Computer Science, University of Trento, Italy}
\affil[6]{\small Orbital EOS - Earth Observation Solutions, Spain}

\affil[*]{\small Corresponding author: \texttt{igor.atake@cmcc.it}}

\date{}  

\begin{document}
\maketitle

\begin{abstract}
Accurate simulations of oil spill trajectories are essential for supporting practitioners’ response and mitigating environmental and socioeconomic impacts. Numerical models, such as MEDSLIK-II, simulate advection, dispersion, and transformation processes of oil particles. However, simulations heavily rely on accurate parameter tuning, still based on expert knowledge and manual calibration. To overcome these limitations, we integrate the MEDSLIK-II numerical oil spill model with a Bayesian optimization framework to iteratively estimate the best physical parameter configuration that yields simulation closer to satellite observations of the slick. We focus on key parameters, such as horizontal diffusivity and drift factor, maximizing the Fraction Skill Score (FSS) as a measure of spatio-temporal overlap between simulated and observed oil distributions. We validate the framework for the Baniyas oil incident that occurred in Syria between August 23 and September 4, 2021, which released over 12,000 m$^3$ of oil. We show that, on average, the proposed approach systematically improves the FSS from 5.82\% to 11.07\% compared to control simulations initialized with default parameters. The optimization results in consistent improvement across multiple time steps, particularly during periods of increased drift variability, demonstrating the robustness of our method in dynamic environmental conditions.
\end{abstract}




\section{Introduction}

Oil spill pollution has historically been linked to rare but catastrophic events like \textit{Deepwater Horizon}, \textit{Prestige}, \textit{Erika}, and \textit{Jyieh} (\cite{boufadel2023, balseiro2003, daniel2001, coppini2011}). Since the 1970s, organizations such as the International Tanker Owners Pollution Federation Limited and the Regional Marine Pollution Emergency Response Centre for the Mediterranean Sea have documented accidental spills, creating a valuable long-term dataset (\cite{polinov2021}). However, limitations in satellite revisit times and coverage hinder real-time detection, especially during initial spill formation, critical for mitigating damage. Thus, numerical models remain essential for simulating oil spill dynamics and assessing broader consequences (\cite{mills2016, jones2017}). While oil spill modeling has advanced (\cite{huang1983, spaulding1988, spaulding2017, reed1999, keramea2021}), drift simulation methods, predominantly Eulerian or Lagrangian, have changed little (\cite{barker2020}). The Lagrangian approach tracks oil as discrete elements, each representing a droplet size class that moves according to physical processes (\cite{barker2020}). The DeepWater Horizon spill highlighted key limitations in existing models, including gaps in observed oil spill in the water column and uncertainties in wind, wave, and current inputs (\cite{spaulding2017, boufadel2020, boufadel2023, keramea2023b}). However, it also drove the introduction of some advancements, such as Lagrangian coherent structures to capture features like the “tiger tail” (\cite{peacock2013}). Despite progress, operational spill modeling still struggles with emergency response demands requiring fast, accurate trajectory predictions (\cite{barker2020}). Models' effectiveness depends on accurate input and ambient parameters, often derived from theoretical models or expert judgment, introducing uncertainty in representing complex ocean and weather dynamics (\cite{wunsch1998, deDominicis2012, deDominicis2013b, caron2019}). Enhancing model calibration is crucial for reliable forecasts and spill response. With expanding remote sensing capabilities (\cite{leifer2012, fingas2013, fingas2018, fingas2025}), more events can be analyzed to refine numerical models.

Traditional calibration methods, such as manual tuning or exhaustive search strategies like Grid Search (\cite{belete2021}) and Random Search (\cite{andradóttir2006}), are computationally expensive and inefficient in high-dimensional parameter spaces. Bayesian optimization provides a powerful alternative by efficiently optimizing complex, expensive-to-evaluate functions (\cite{frazier2018}). Unlike traditional methods, Bayesian optimization leverages previous evaluations to statistically prioritize promising parameter regions, accelerating convergence while minimizing computational cost. This approach has proven effective in diverse fields, including materials and drug design (\cite{negoescu2011, frazier2016, packwood2017, zhang2020}), reinforcement learning (\cite{lizotte2010, brochu2009}), and hyperparameter tuning in deep learning (\cite{snoek2012, snoek2015, ranjit2019}). In this work, we integrate the Bayesian optimization framework with the MEDSLIK-II oil spill model (\cite{deDominicis2013a}) to estimate key physical parameters from initial oil slick observations. By refining physical parameters based on observations, this approach (i) enhances model accuracy and reduces modelling biases in the input forcing, and (ii) improves both rapid-response forecasting and short-term (up to 3 days) environmental assessments, making the approach applicable in an operational context. The rest of the article is structured as follows: Section \ref{sec2} details the data and the methodology used to develop and evaluate the proposed solution, Section \ref{sec3} presents results, and Section \ref{sec4} discusses implications, limitations, and future research directions.
%
%
%
%
%
%
\section{Materials and Methods}\label{sec2}

\subsection{Study Area and Event}\label{subsec2.1}
We focus on the Baniyas Thermal Station spill in Syria, which occurred in August 2021 due to years of neglected maintenance following the civil war in the country. The accident released over 12,000 m\(^3\) of oil between August 23 and September 4, when the leakage was finally contained, with oil slicks being identified as far north as Cyprus (\cite{abou2024}). Throughout this period, satellite observations tracked the oil slick’s spread in the Levantine Basin, with Optical and Synthetic Aperture Radar (SAR) imagery capturing its evolution. These observations were provided by OrbitalEOS\footnote{https://www.orbitaleos.com/} in the context of the iMagine\footnote{https://www.imagine-ai.eu/case-study/oil-spill-detection-oil-spill-detection-from-satellite-images} project, which collected, analyzed, and published extensive oil spill imagery as open data on Zenodo\footnote{https://zenodo.org/records/11354663} (\cite{ferrer2024}).

We selected this case study since a considerable number of satellite detections are available along the entire oil slick trajectory, contrary to other oil spill events. However, we focused on the first three days of the event (up to August 27, 2021) due to the extension of the oil slick and the complex mesoscale dynamics that created “tiger tail” features. Such conditions are very difficult to simulate with traditional Lagrangian oil spill modeling because of the submesoscale and mesoscale interactions (\cite{olascoaga2012}).

\begin{figure}[ht]
    \centering
    \includegraphics[width=0.8\textwidth]{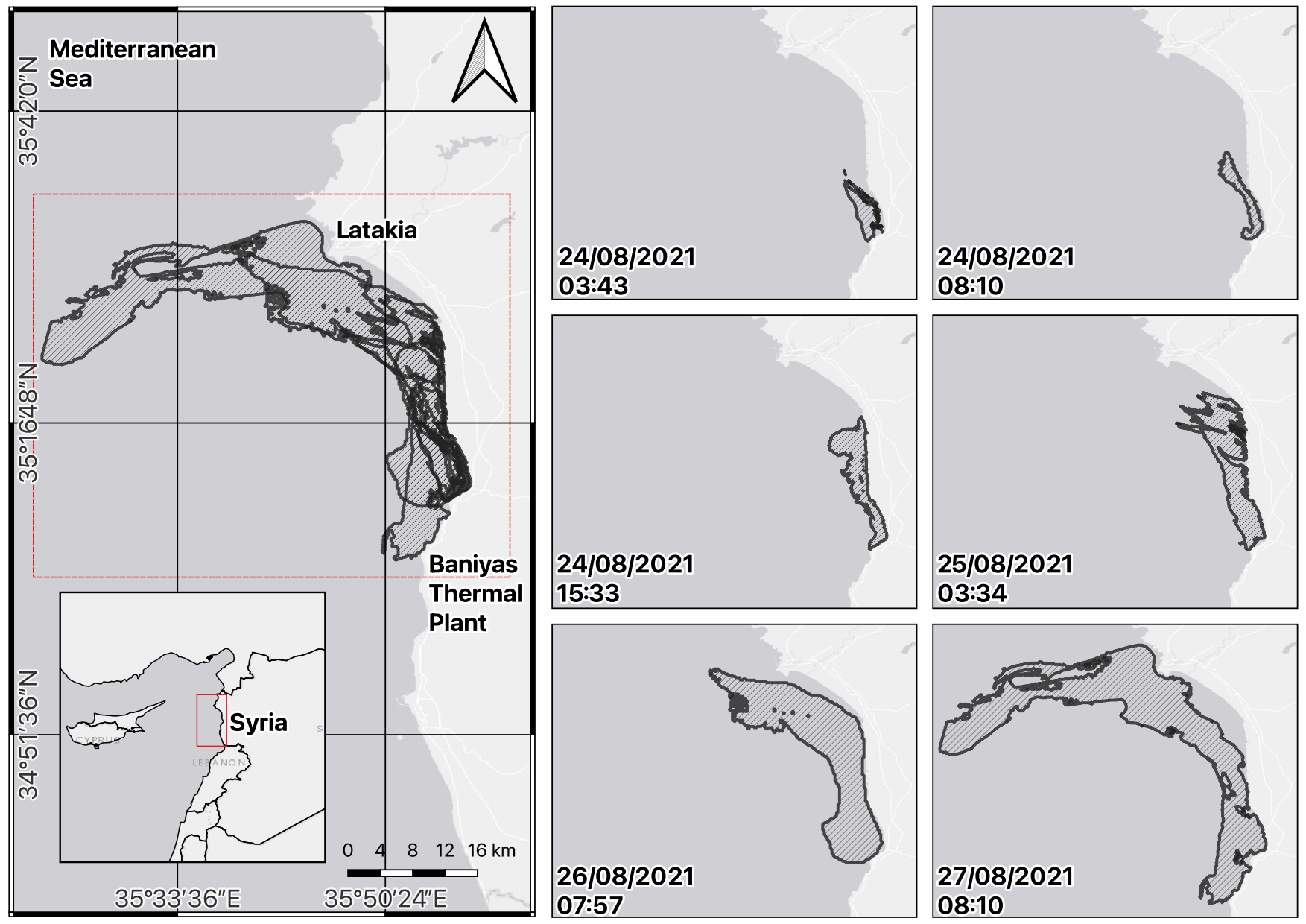}
    \caption{Oil slicks selected from August 24$^{th}$ to 27$^{th}$, 2021 targeting the Baniyas oil spill accident in Syria. Each panel reports consecutive stages of the oil slick according to satellite retrievals and revisit times.}
    \label{fig:figure1}
\end{figure}

Considering the data and our premises, Figure \ref{fig:figure1} presents all the identified and extracted oil slicks over the time period of this study.

The study area is located within the Levantine basin. The mesoscale currents pattern in the vicinity of the accident is governed by a relation between anticyclonic and cyclonic eddies in the Latakia region. Oceanographic surveys and models have contributed to the understanding of these features over time, which lead to a northward transport in the easternmost part, in processes that involve the Libyan-Egyptian Current, the Asia Minor Current, and the persistent and temporary eddies in the region (\cite{robinson1987, hecht1988, zodiatis2023, keramea2023a}).

\begin{figure}[htbp]
    \centering
    \includegraphics[width=\textwidth]{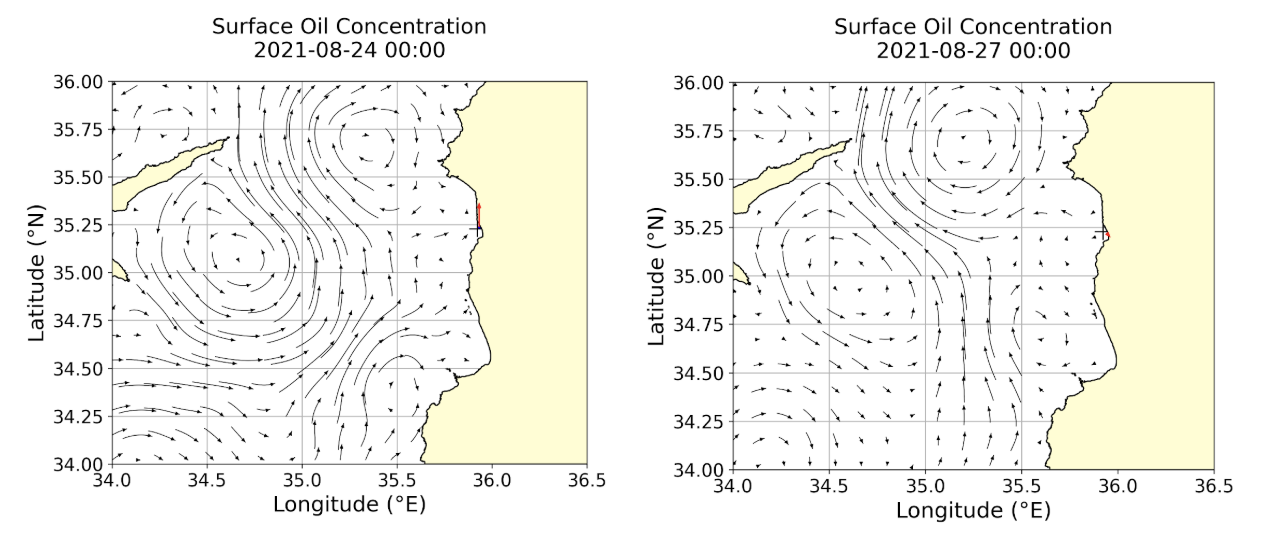}
    \caption{Surface currents from the Copernicus Marine reanalysis product in the Levantine Basin during the study period. The left panel shows the initial conditions on August 24, 2021, at 00:00, while the right panel displays the final conditions on August 27, 2021, at 12:00.}
    \label{fig:figure2}
\end{figure}

In Figure~\ref{fig:figure2}, the pattern of sea surface currents, from the Mediterranean Sea reanalysis (\cite{escudier2020}), is shown at 24-08-2021 00:00 and 28-08-2021 12:00, representing the initial and final conditions of the proposed experiments. The anticyclonic eddy in the south and the cyclonic eddy in the north are persistent features during the period, while their position and consequently the intensity and shape of the northward currents are shifted horizontally in the western direction.

\subsection{Data}\label{subsec2.2}
\begin{enumerate}
    \item \textbf{Oil Spill Imagery} \\
    The dataset of segmented observations consists of oil slick boundary shapefiles extracted from the raw satellite imagery from 2019 to 2022. The entire dataset (\cite{ferrer2024}) is available through a data platform hosted at the University of Trento and developed in the context of the iMagine project. Such a platform includes services exposing oil spill data via HTTP (i.e., THREDDS\footnote{http://thredds.imagine.disi.unitn.it/thredds/catalog/catalog.html}) and a Graphical User Interface (GUI) providing data access, search \& discovery as well as advanced data visualization capabilities. Due to page limitations, the GUI is beyond the scope of this paper and will be presented in a future work.
    Considering the sparse nature of oil spill observations (\cite{shaban2021}) and the requirement of at least two observations for analysis, we focused on the most reliable and consistent data available to evaluate the Bayesian optimization framework in the study area (Figure~\ref{fig:figure1}). Table~\ref{tab:table1} reports the slicks selected from the dataset.
        
    \begin{table}[!ht]
    \caption{Remotely sensed oil slick observations provided by Orbital EOS.\label{tab:table1}}
    \begin{threeparttable}
    \begin{tabular*}{\columnwidth}{@{\extracolsep\fill}cccc@{\extracolsep\fill}}
    \toprule
    \textbf{Observation ID} & \textbf{Passing time (UTC)} & \textbf{Satellite} & \textbf{Sensor type} \\
    \midrule
    O1 & 24-08-2021 03:43 & Sentinel-1 & SAR \\
    O2 & 24-08-2021 08:10 & Sentinel-3 & SAR \\
    O3 & 24-08-2021 15:33 & Sentinel-1 & SAR \\
    O4 & 25-08-2021 03:34 & Sentinel-1 & SAR \\
    O5 & 26-08-2021 07:57 & Sentinel-3 & Optical \\
    O6 & 27-08-2021 08:10 & Landsat 8 & SAR \\
    \bottomrule
    \end{tabular*}
    \end{threeparttable}
    \end{table}

    \item \textbf{Meteo-oceanographic Input} \\
    Meteo-oceanographic data provides the necessary environmental forcing for oil spill models. All required datasets were retrieved via web services, corresponding to the event timelines, and extracted within a 2-3 degree radius around the centroid of the first observation. Marine currents and sea surface temperature (SST) data were sourced from the European Copernicus Marine Service and the Copernicus Climate Change Service, from the Mediterranean Sea reanalysis dataset (\cite{escudier2020}). This dataset offers surface and subsurface currents and SST at a 1/24° resolution ($\sim$4.5 km) grid. Wind data at 10 m were obtained from the ERA5 reanalysis on single levels dataset (\cite{copernicus2024}), provided at a 1/4° resolution ($\sim$25×25 km grid).
    
    \item \textbf{Bathymetry and Coastline} \\
    Simulations also require bathymetry and coastline data to account for potential shoreline impacts. In the absence of high-resolution local sources, global datasets are adopted. The General Bathymetric Chart of the Oceans 2024 (\cite{mayer2018}) provides sub-ice bathymetry. The Global Self-consistent, Hierarchical, High-resolution Geography Database (\cite{wessel1996}) is used at its highest resolution (10–100 m segments) for coastline representation.
\end{enumerate}

\subsection{MEDSLIK-II oil spill model}\label{subsec2.3}
The Lagrangian oil spill model MEDSLIK-II (\cite{deDominicis2013a, deDominicis2013b}) is an open-source community model that has been used for tracking oil spills in the Mediterranean Sea for nearly 15 years. It discretizes oil slicks into parcels, simulating advection-diffusion and oil weathering processes. Oil transport is driven by water currents, waves, wind, and turbulent fluctuations, modeled via a random walk scheme. MEDSLIK-II includes a proper representation of high-frequency currents and wind fields in the advective components of the Lagrangian trajectory model, the Stokes drift velocity, and the coupling with the remote-sensing data related to observed oil slicks. The governing equation to track oil uses the advection-diffusion equation and is presented in equation \ref{eq:eq2.1}:

\begin{equation}
    \frac{\partial c}{\partial t} + {u} \cdot \nabla c = \nabla (\vec{k} \nabla c) + \sum_{j=1}^{M}r_j(\vec{x},C, t)
    \tag{2.1}\label{eq:eq2.1}
\end{equation}

The term $\frac{\partial c}{\partial t}$ represents the rate of change of concentration over time, with advection captured by the term ${u} \cdot \nabla C$, where ${u}$ is the ambient velocity field. The term $\nabla (\vec{k} \nabla c)$ represents spatially varying diffusivity, and the last term shows the sum of reaction terms accounting for physical-chemical processes such as spreading, evaporation, emulsification, dissolution, photo-oxidation, biodegradation, and sedimentation and adhesion to the coast (\cite{deDominicis2013a}). \\
\cite{deDominicis2013b} analyzed the sensitivity of the model physical parameters to varying inputs, such as wind drift and wind angle, through five experiments. Their findings highlighted the critical role of high-fidelity and high-resolution currents in improving forecast accuracy over a 1 to 2.5-day time-frame. The study also suggested the use of ensemble simulations with uncertain physical parameters to enhance predictive performance over deterministic approaches. Building on this work, we extend the analysis to two selected physical parameters, specifically parametrization coefficients within the processes represented in the formulations. In this work we use the MEDSLIK-II model version 2.01, available as open source at the model website\footnote{https://www.medslik-ii.org/index.html}. The two parametrizations we aim to optimize are:

\begin{enumerate}
    \item \textbf{Horizontal diffusivity} \\
    \begin{equation}
        dx^{\prime}_{k}(t) = Z_{1}\sqrt{2K_{x} dt} = \left[ 2r -1\right] \sqrt{6K_{h}dt}
    \tag{2.2}\label{eq:eq2.2}
    \end{equation}
    \begin{equation}
        dy^{\prime}_{k}(t) = Z_{2}\sqrt{2K_{y} dt} = \left[ 2r -1\right] \sqrt{6K_{h}dt}
    \tag{2.3}\label{eq:eq2.3}
    \end{equation}
    
    where $r$ is a random variable uniformly distributed in $[0, 1]$, and $D$ is the horizontal diffusivity coefficient, varying in the order of magnitude of 1-100 m$^2$/s (\cite{deDominicis2012, deDominicis2013a}).

    \item \textbf{Wind drift and angle} \\
    \begin{equation}
        U_{W} = \alpha (W_{x}\cos(\beta) + W_{y}\sin(\beta))
    \tag{2.4}\label{eq:eq2.4}
    \end{equation}
   \begin{equation}
        V_{W} = \alpha (-W_{x}\sin(\beta) + W_{y}\cos(\beta))
    \tag{2.5}\label{eq:eq2.5}
    \end{equation}

    Where $U$ and $V$ represent the wind zonal and meridional components at 10 m, while $\Delta v$ and $\theta$ correspond to the wind drift and wind angle, respectively. The MEDSLIK-II model typically assumes a wind drift of 3\% and wind angle between 0.0-25.0$^{\circ}$ (\cite{alRabeh2000}). In this study, we set the wind angle to $0.0^{\circ}$ and kept it constant over simulations, since its effect on correcting local Ekman currents is difficult to validate and could negatively impact the results (\cite{deDominicis2013b}).
\end{enumerate}   

\subsection{Bayesian Optimization}\label{subsec2.4}
Bayesian optimization (\cite{kushner1964}; \cite{zhilinskas1976}; \cite{mockus1975}; \cite{mockus1989}; \cite{jones1998}) is a class of data-driven optimization methods designed for black-box derivative-free global optimization problems. The goal is to maximize an objective function $f(x)$,

\begin{equation}
    x^{*} = \arg\max_{x \in \mathcal{X}} f(x)
    \tag{2.6}\label{eq:eq2.6}
\end{equation}

where $x \in \mathcal{X}$ is the physical parameter vector and $\mathcal{X} \subseteq \mathbb{R}^n$ denotes the \textit{n}-dimensional parameter space (\cite{frazier2018}). This approach is particularly advantageous when function evaluations are computationally expensive due to the high dimensionality of the parameter space, the computational cost of resource-intensive simulations, or when the analytical form of $f(x)$ is unavailable or unknown. At its core, Bayesian optimization relies on a probabilistic surrogate model, $\hat{f}(x)$, to approximate the behavior of the objective function. A Gaussian Process (GP) (\cite{rasmussen2006, naveiro2024}) is commonly used as the surrogate model due to its ability to flexibly represent a wide range of functions while quantifying uncertainty and providing an analytical posterior distribution, $\hat{f}(x) | D$ , where $D={((x_i, f(x_i)) | i = 1, ..., m)}$ is the set of previously observed evaluations. The posterior distribution enables probabilistic predictions of the objective function’s value at unobserved points. Using the posterior distribution, Bayesian optimization selects the next evaluation point based on an acquisition function. This function is designed to balance (i) sampling on regions of high uncertainty in the parameter space (exploration), and (ii) sampling on regions likely to yield high objective values based on current knowledge (exploitation). A common choice for the acquisition function is the Upper Confidence Bound (UCB), defined as:

\begin{equation}
    UCB(x) = \mu(x) + k \cdot \sigma(x)
    \tag{2.7}\label{eq:eq2.7}
\end{equation}

where $\mu(x)$ and $\sigma(x)$ are the mean and standard deviation of the posterior distribution at $x$ and $k > 0$ is a parameter controlling the trade-off between exploration and exploitation (\cite{carpentier2015}). Bayesian optimization iteratively refines the surrogate model and acquisition function by incorporating new samples. At each iteration, the parameter vector is sampled as:

\begin{equation}
    \hat{x} = \arg\max_{x \in \mathcal{X}} UCB(x)
    \tag{2.8}\label{eq:eq2.8}
\end{equation}

The objective function is evaluated at $\hat{x}$, and the surrogate model is updated with the new sample point $(\hat{x}, f(\hat{x}))$. This process continues until predefined convergence criteria, such as a maximum number of iterations or a threshold improvement in $f(x)$, are satisfied.

\subsection{Coupled Bayesian Optimization \textemdash MEDSLIK-II framework}\label{subsec2.5}
In this study, we aim to identify the optimal parameter configuration of the MEDSLIK-II model that yields simulations most closely aligned with the spatio-temporal distribution of observed oil slicks, as defined in Section \ref{subsec2.1}. These simulations depend on the specific configuration of a parameter vector. We define the tunable parameter vector as $x = \left[ K_h, \alpha \right]^T \in \mathcal{X}$, where $ \mathcal{X} \subseteq \mathbb{R}^n$ is the parameter space, $K_h$ represents the horizontal diffusivity, and $\alpha$ denotes the wind drift. As also stated in Section \ref{subsec2.3}, the wind angle parameter $\beta$ (equations \ref{eq:eq2.4} and \ref{eq:eq2.5}) is excluded from optimization and it is set to $0.0^\circ$ in all the experiments (\cite{deDominicis2013b}). The ranges for the tunable parameters are defined as $K_h \in [0.0, 20.0]$ m$^2$/s and $\alpha \in [0.0, 0.05]$ m/s after an initial sensitivity test, but also respecting the boundaries found in the literature (\cite{deDominicis2012}).

\begin{figure}[htbp]
    \centering
    \includegraphics[width=\textwidth]{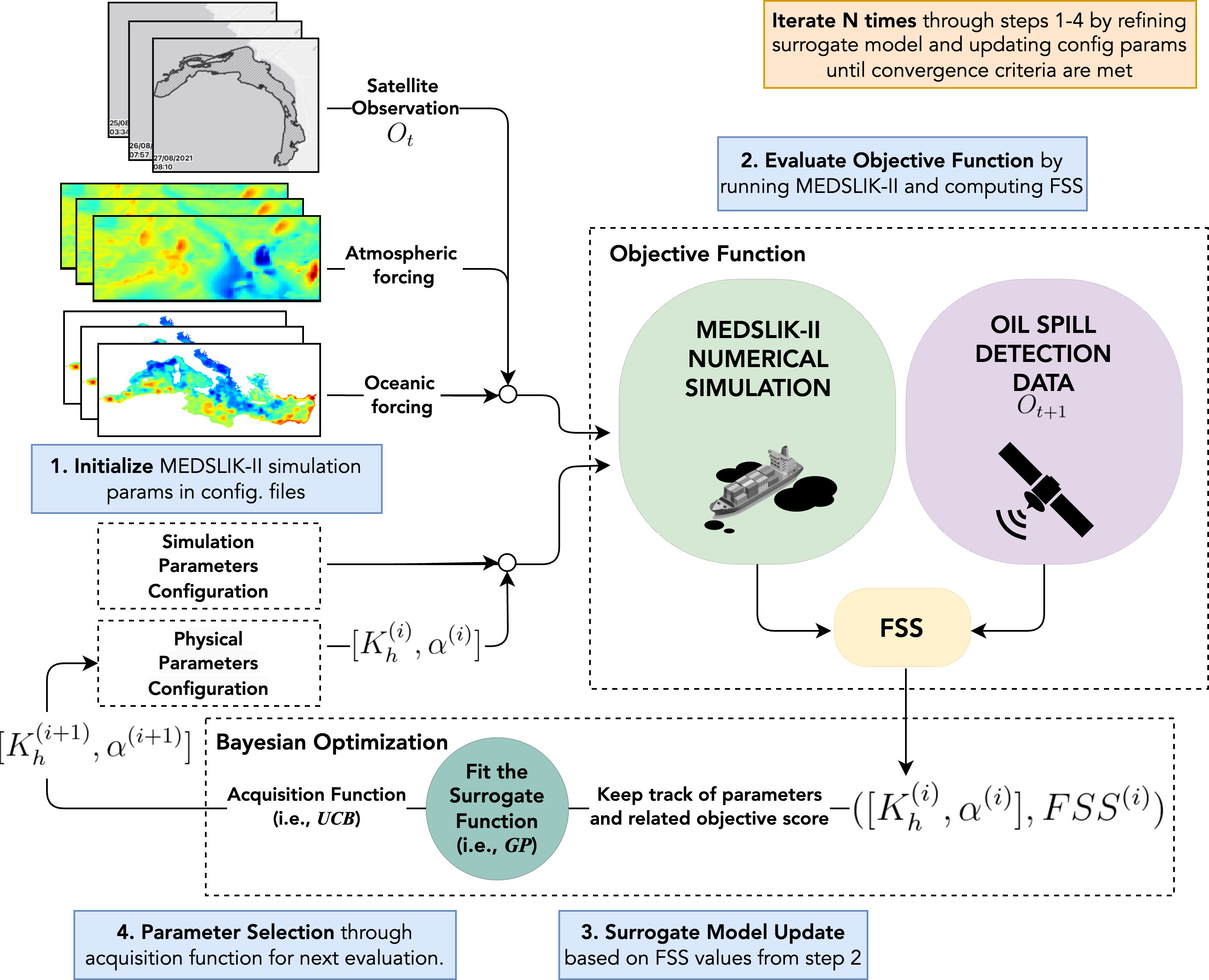}
    \caption{Overview of the Bayesian optimization Workflow designed for improving the accuracy of the MEDSLIK-II oil spill model simulations. The workflow iteratively refines the simulation configuration parameters by evaluating the objective function and optimizing the acquisition function until a convergence criteria is met. Here, we denote the iteration number with the superscript (\textit{i}), for the sake of clarity.}
    \label{fig:figure3}
\end{figure}

To evaluate the agreement between simulated and observed oil slick distributions, we adopt the Fractions Skill Score (FSS) as the optimization metric (\cite{roberts2008}; \cite{robertsandlean2008}), which is further explained in Section \ref{subsec2.6}. It represents the fractional coverage of simulated and observed oil particles within a specified spatial neighborhood, ranging from 0 (no skill) to 1 (perfect skill), with higher values indicating better spatial alignment. Hence, the Bayesian optimization task is formulated to maximize this score:

\begin{equation}
    x^{*} = \arg\!\max_{x_{i} \in \mathcal{X},  i = 1, \dots, N} FSS(\mathcal{S}_{x_i}, \mathcal{O}_{t+1})
    \tag{2.9}\label{eq:eq2.9}
\end{equation}

where $FSS(S_{x_{i}}, \mathcal{O}_{t+1})$ serves as the objective function to be maximized. Here, $S_{x_{i}}$ represents the MEDSLIK-II simulation initialized at the observation $O_t$, based on the corresponding atmospheric and oceanic forcing at that time, along with the physical parameter vector $x_i$. The term $\mathcal{O}_{t+1}$ denotes the subsequent observed oil slick data used for comparison, while $x^*$ represents the optimal parameter vector obtained upon convergence. Beyond physical parameters, the MEDSLIK-II simulation is also configured with simulation parameters that include oil-specific information. Notably, the same oil type—Safaniya, an Arabian heavy oil with an American Petroleum Institute gravity of 27.9—was used for all simulations. To maintain consistency, we did not alter values related to oil chemistry, instead focusing on the trajectory of the oil spill while keeping its chemical processes fixed. In the experiments, we set $t \in [{1,...,5}]$, aligning with the available observations (Table \ref{tab:table1}), and limit the maximum number of iterations in the Bayesian optimization framework to $N=100$. The process requires the evaluation of an objective function, using a "one-step" assumption, which quantifies the accuracy of the MEDSLIK-II model's output relative to target observation(s). To approximate the behavior of $FSS(S_{x_{i}}, \mathcal{O}_{t+1})$, a Gaussian Process (GP) is employed as a surrogate model. At each iteration $i$, the MEDSLIK-II model is initialized with the current parameter vector $x_i$, and the corresponding $FSS(S_{x_{i}}, \mathcal{O}_{t+1})$ is computed. For the first iteration $(i=1)$, the parameter vector is initialized to default values, $K_h = 2.0$ and $\alpha=0.0$ (\cite{deDominicis2013a}). The parameter vector for the subsequent iteration $i+1$ is determined by maximizing the acquisition function $UCB$ as in equation \ref{eq:eq2.7}, with $k=2.576$ (\cite{jones1998}; \cite{srinivas2010}; \cite{frazier2018}). After evaluating the objective function at $x_{i+1}$, the GP is updated with the new sample $(x_{i+1}, FSS(S_{x_{i}}, \mathcal{O}_{t+1}))$. We repeat this process until we reach $N$ for a given observation. The overall workflow is presented in Figure ~\ref{fig:figure3}. For the implementation of Bayesian optimization, we used the BayesOpt python library\footnote{https://github.com/bayesian-optimization/BayesianOptimization}, which was customized to meet the specific needs of this study.

\subsection{Experimental Settings}\label{subsec2.6}
This section outlines the experimental settings used to evaluate the skills of the proposed approach across six sequential observations, reported in Table~\ref{tab:table1}. To compare MEDSLIK-II simulations with observations, we performed a series of experiments forcing MEDSLIK-II with the default physical parameters and with the ones estimated through Bayesian optimization. Specifically, we consider a control phase in which the MEDSLIK-II model is initialized with default physical parameters and a validation phase in which the Bayesian optimization is employed to sample the best simulation parameter configuration based on available observations. The control phase mirrors a typical scenario in which default physical parameters are provided by the model configuration and remain fixed throughout all simulations.

For the validation phase, we considered two approaches, which represent a trade-off between computational efficiency and accuracy: 

\begin{enumerate}[i.]
    \item \textbf{One-step Optimization (OSO)}: observation O1 is used as the initial condition for MEDSLIK-II. Bayesian optimization is then performed using observation O2 to determine the optimal physical parameters. These parameters are subsequently applied in the numerical model to simulate the oil slick trajectory for all future time steps without further adjustments;
    \item \textbf{Multi-step Optimization (MSO)}: Bayesian optimization is applied iteratively between consecutive observations. At each step, the optimal parameters derived from the comparison between two successive observations (e.g., O1 → O2) are used to simulate the trajectory up to the next observation (e.g., O2 → O3). This process continues iteratively, refining the parameters at each step.
\end{enumerate}

\begin{figure}[htbp]
    \centering
    \includegraphics[width=\textwidth]{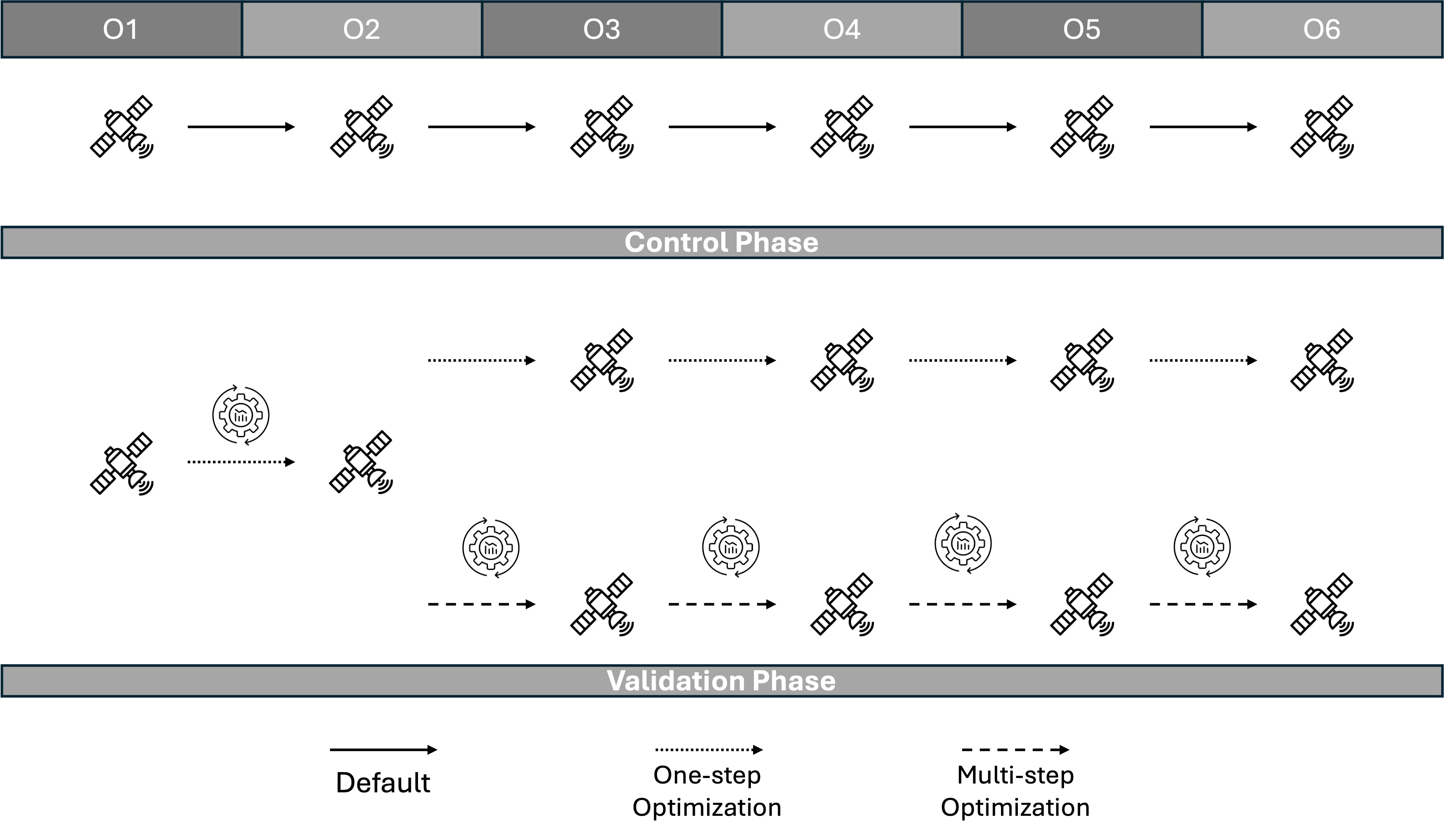}
    \caption{Experimental settings. The control phase consists of running MEDSLIK-II simulations with standard physical parameters. The validation phase is divided into two approaches: One-step Optimization and Multi-step Optimization. In the former, optimization is applied only from observation O1 to O2, with the resulting physical parameters used for all subsequent simulations. In the latter, optimization is performed at each observation step, to capture potential variations in model physical parameters.}
    \label{fig:figure4}
\end{figure}

While the former prioritizes computational efficiency for operational use, the latter maximizes the accuracy through iterative recalibration of model physical parameters, enhancing flexibility and adaptability by continuously updating conditions. The overview of the experimental settings is depicted in Figure~\ref{fig:figure4}.

To assess the accuracy of the simulated oil slick with respect to observations, we adopt two different metrics: the Fractions Skill Score (FSS), which is also used as the objective function of the Bayesian optimization framework, and the Centroid Skill Score (CSS). The chosen metrics are described below:

\begin{enumerate}
    \item \textbf{Fractions Skill Score (FSS)}\\
    The FSS was originally introduced in (\cite{robertsandlean2008}) for comparing observed and modeled rainfall accumulation. Later, it was also applied in oil spill modeling, demonstrating the suitability for operational oil spill tracking purposes (\cite{simecekBeatty2021}). In this work, the FSS measures the skill of oil spill simulations by comparing them against a ground truth (i.e., satellite observations), after the interpolation of both data onto a common $150\,\text{m} \times 150\,\text{m}$ grid over the area of study. Unlike the traditional overlay method, this metric considers not only the common area between simulation and observation but also the spatial distribution within that area. The FSS is calculated according to the following formula:

    \begin{equation}
    	FSS = 1 - \frac{\sum_{i=1}^{n}(f_i - o_i)^2}{\sum_{i=1}^{n}(f_i^{2} + o_i^{2})}
    \tag{2.10}\label{eq:eq2.10}
    \end{equation}

    where $f_i$ is the forecast fraction, $o_i$ is the observed fraction, and the summation runs over all grid points $i$. The FSS ranges from 0 to 1, where 1 indicates a perfect match between the forecast and observation, and 0 indicates no skill.
    
    \item \textbf{Centroid Skill Score (CSS)}\\  
    The Centroid Skill Score (\cite{liu2011}) provides a complementary assessment by measuring the Lagrangian separation distance between the centroids of observed and simulated oil spill areas normalized by the observed spill size:

    \begin{equation}
    CSS =\begin{cases} 1-\frac{CI}{CI_{thr}} & \text{if CI < $C_{thr}$,} \\0,  & \text{otherwise.} \end{cases}
    \tag{2.11}\label{eq:eq2.11}
    \end{equation}
    
    where $CI=\frac{\Delta{X}}{L_{obs}}$ the Lagrangian separation distance (in \textit{km}) between SAR-derived and modeled oil slick at a given time and $L_{obs}$ is the diagonal length (in \textit{km}) of the bounding box enclosing the observed slick. A CSS of 1 indicates a perfect match while a CSS of 0 means the centroids are distant relative to the observation size. Following \cite{deDominicis2014}, the threshold $C_{thr}$ is set to 1.
\end{enumerate}

\section{Results}\label{sec3}
Table \ref{tab:table2} presents the quantitative results for all tested experiments, showcasing each simulation metric across different settings: control, OSO, and MSO. For each observation, we report the FSS and CSS scores along with their percentage difference over the control setting when applying the OSO and MSO strategies. The specific physical parameters used in each MEDSLIK-II simulation are detailed in Table \ref{tab:tableS1}. The default values come from the standard MEDSLIK-II model, while the OSO parameters were optimized using Bayesian optimization between observations O1 and O2. The MSO parameters, on the other hand, were refined through continuous training over successive observation pairs.

\begin{table}[!ht]
\caption{FSS and CSS results for control, OSO, and MSO settings across the assessed observations. Additionally, we report the percentage difference for both OSO and MSO relative to the control setting, as well as the average values across the different observations and settings.\label{tab:table2}}
\begin{threeparttable}
\begin{tabular*}{\columnwidth}{@{\extracolsep{\fill}}lcccccc@{\extracolsep{\fill}}}
\toprule
\textbf{Obs} & \textbf{Metric} & \textbf{Control} & \textbf{OSO} & \textbf{OSO/Control \%} & \textbf{MSO} & \textbf{MSO/Control \%} \\
\midrule
\textbf{O2} & FSS & 0.63 & 0.66 & 4.76 & - & - \\
                          & CSS & 0.86 & 0.94 & 9.30 & - & - \\
\midrule
\textbf{O3} & FSS & 0.57 & 0.65 & 14.03 & 0.67 & 17.54 \\
                          & CSS & 0.82 & 0.96 & 17.07 & 0.96 & 17.07 \\
\midrule
\textbf{O4} & FSS & 0.59 & 0.59 & 0.0 & 0.67 & 13.56 \\
                          & CSS & 0.96 & 0.97 & 1.04 & 0.98 & 2.08 \\
\midrule                          
\textbf{O5} & FSS & 0.60 & 0.64 & 6.6 & 0.69 & 7.81 \\
                          & CSS & 0.84 & 0.87 & 3.57 & 0.84 & 0.0 \\
\midrule                          
\textbf{O6} & FSS & 0.54 & 0.56 & 3.7 & 0.59 & 5.36 \\
                          & CSS & 0.87 & 0.86 & -1.15 & 0.85 & -2.3 \\
\hline
\hline
\textbf{Avg.} & FSS & 0.59 & 0.62 & 5.82 & 0.66 & 11.07 \\
                          & CSS & 0.87 & 0.92 & 5.97 & 0.91 & 4.21 \\
\bottomrule
\end{tabular*}
\end{threeparttable}
\end{table}

\begin{figure}[htbp]
    \centering
    \includegraphics[width=\textwidth]{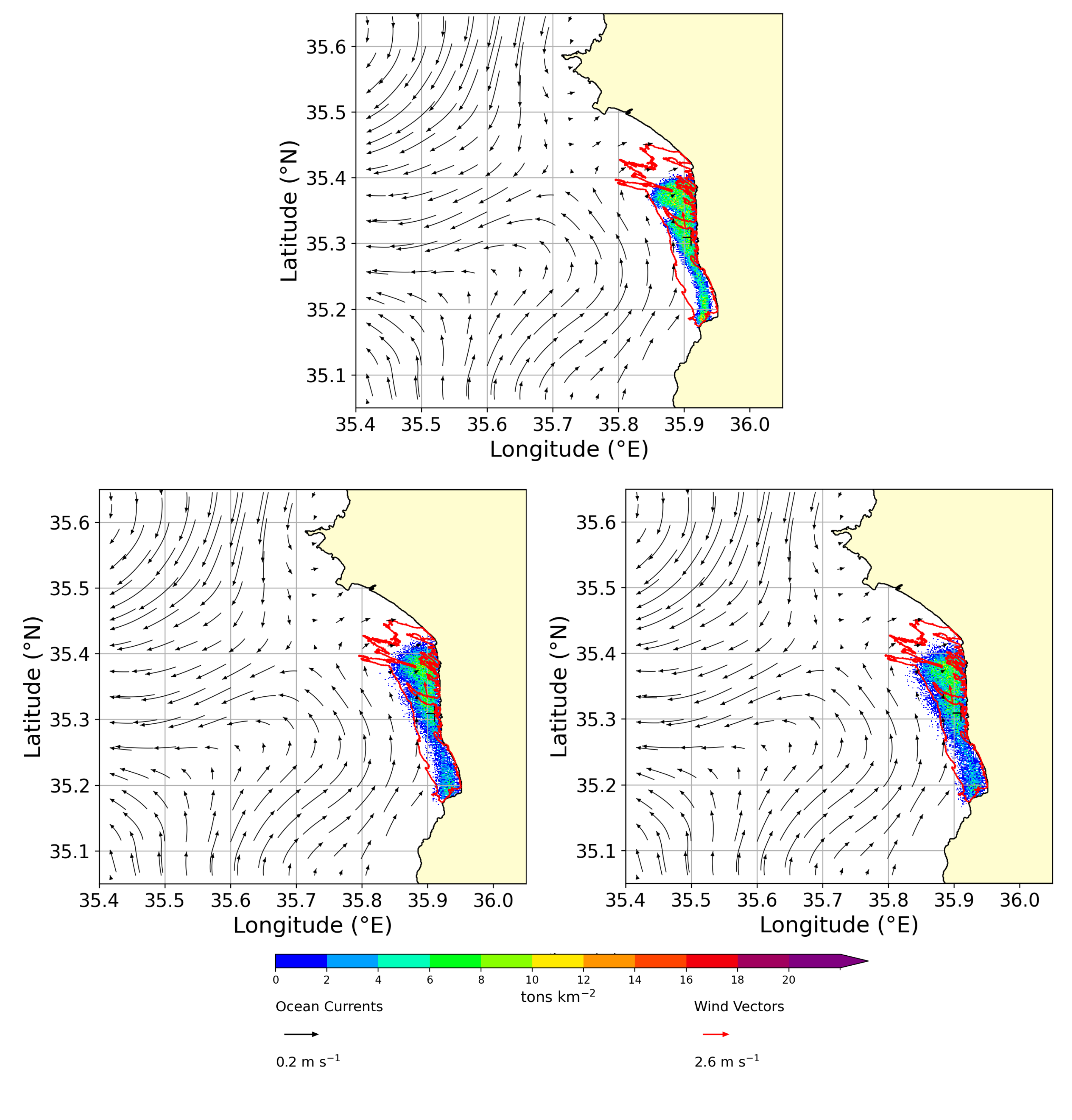}
    \caption{MEDSLIK-II surface oil concentrations starting from O3 (24/08/2021 15:33) as initial condition and stopping at O4 (25/08/2021 03:34). The upper panel represents control simulation with default physical parameters. Bottom panel represents the optimized simulation using the OSO (left) and MSO (right) settings.}
    \label{fig:figure5}
\end{figure}

Figure~\ref{fig:figure5} illustrates the results for the control, OSO, and MSO approaches, using observation O3 (see Table \ref{tab:table1}) as the initial condition. Each evaluation metric assesses the accuracy of MEDSLIK-II simulations against satellite observations of the oil slick trajectories, following temporal alignment. Since the observations were provided as shapefiles without oil slick thickness data, we focused on the slick shapes for area comparison. Consequently, we employed the geographical centroid in our analysis, as the center of mass could not be determined from the available data.
Because the optimization process targeted the FSS metric, we consistently observe an improvement from control to OSO and further to MSO. However, this pattern is not always reflected in CSS scores. For example, in observation O3, both FSS and CSS reached their highest values in MSO.
A closer look at observation O4 in Figure \ref{fig:figure5} reveals a clear skill enhancement from the control setting to OSO and MSO across all metrics. In the control run, the simulated oil slick appears narrower, remaining closer to the coast and failing to move as far north as in the OSO and MSO experiments. Comparing OSO and MSO, we see that MSO achieves a slightly better match, particularly in the southern region of the observed slick. These differences can largely be attributed to variations in horizontal diffusivity. In the control setting (horizontal diffusivity set to $2.0$ m$^2$/s), the slick is advected with minimal spreading. In contrast, the optimized horizontal diffusivity values for OSO ($10.26$ m$^2$/s) and MSO ($12.68$ m$^2$/s) allow the slick to spread more effectively, covering a larger portion of the observed area and improving accuracy.

\begin{figure}[htbp]
    \centering
    \includegraphics[width=\textwidth]{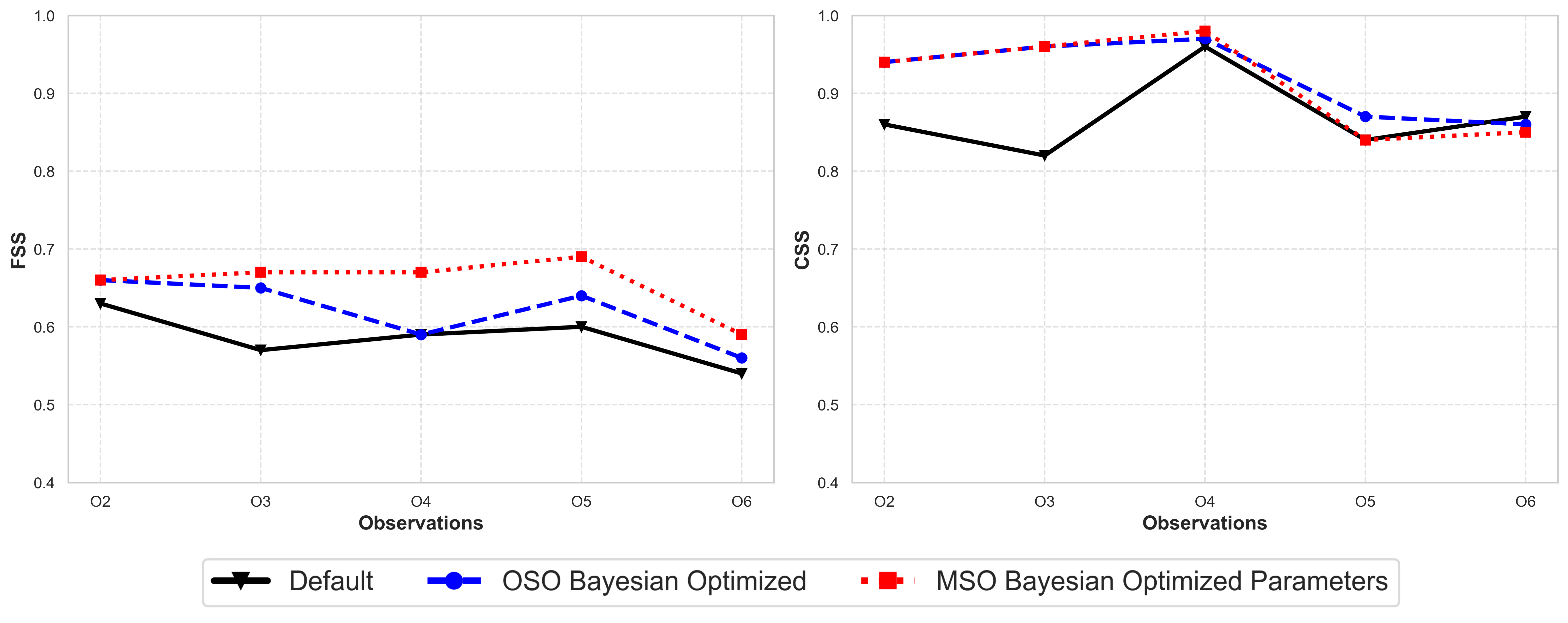}
    \caption{Comparison of the FSS (left panel) and CSS (right panel) metrics between the simulation using optimal parameters identified through Bayesian optimization and the simulation with default MEDSLIK-II parameters. The x-axis reports each observation (see Table~\ref{tab:table1}), while the y-axis reports the values of each metric, both bounded in $[0,1]$ (higher values are better). Here, we do not report observation O1 which is instead used for initializing the MEDSLIK-II simulations in all the settings.}
    \label{fig:figure6}
\end{figure}

Our simulations, initialized with the Bayesian optimization estimated parameters, revealed substantial improvements across both settings. The FSS metric in the OSO configuration, outperformed the control setting by ~5.82\% on average, while the MSO configuration improved even further by ~11.07\%. Regarding the CSS metric, the optimization also enhanced the average accuracy by 5.97\% for the OSO strategy and 4.21\% for MSO, further underscoring the advantages of Bayesian optimization. However, despite its superior accuracy, MSO comes with added complexity, making its implementation in operational settings more challenging, as previously discussed. Nevertheless, when comparing OSO and MSO, it is evident that the latter consistently outperforms both the control and OSO settings for FSS, demonstrating a sustained improvement in accuracy (see Figure~\ref{fig:figure6}). However, this comes at the expense of increased computational cost.

When comparing the visual characteristics of the oil spill in Figure~\ref{fig:figure5} with Figures \ref{fig:figureS1}, \ref{fig:figureS2}, \ref{fig:figureS3}, and \ref{fig:figureS4} in the Appendix A, it becomes clear that the modeled results better match the observed oil slick extent in the first three simulations. However, as reported in Table \ref{tab:table2}, the percentage differences for both OSO and MSO relative to the control setting are negative for observation O6 (-1.15\% and -2.3\%, respectively). Although the magnitude of these values is small, this highlights the importance of assessing both metrics together, as they are able to capture different features and characteristics of the oil slick.

\section{Discussion}\label{sec4}
\subsection{Considerations around metrics and data used}\label{subsec4.1}
The results presented in Section \ref{sec3} indicate that simulation accuracy, measured by the optimized metric (FSS), consistently improved when the Bayesian optimization workflow was applied. Both the MSO and OSO approaches led to higher FSS values compared to the control simulation setup, with MSO generally outperforming OSO. However, a few exceptions were observed when considering the CSS metric. This does not necessarily imply poorer results, as area-based metrics like FSS and centroid-based metrics such as CSS complement each other in assessing oil spill modeling accuracy. While CSS primarily indicates the accuracy of the oil spill's location, FSS evaluates how well the model captures spreading and weathering processes (\cite{dearden2021}). Additionally, FSS scores are typically lower than CSS scores, as accurately matching the precise area and shape of an oil slick is inherently more challenging than achieving strong performance on geographical centroids, even after optimization.
Although the number of variables was limited in all experiments, other inputs could have significantly influenced the results. For instance, additional MEDSLIK-II parameters—such as the spreading rate of thick and thin slicks or evaporation parameters—different environmental forcings, including regional or local circulation models tailored to the period of interest, or more detailed observations of the oil slick characteristics (e.g., oil thickness) could have been considered.  
As shown in the surface current data in Figure \ref{fig:figure2}, the primary oceanographic features are two eddies that drive a northward current pattern. While these features are captured in the forcing data, the accident occurred along the coastline, far from these mesoscale structures. Additionally, the reanalysis data used for coastal conditions lacks the resolution necessary to accurately represent coastal currents at the scale at which circulation models are generated.
MEDSLIK-II can extrapolate currents from areas with available data to cells where no current values are provided. This extrapolation averages currents in ocean cells until land cells are reached (\cite{deDominicis2012}). However, this method does not ensure that the real circulation features are accurately represented in the extrapolated region. 
The combination of general meteo-oceanographic conditions, data limitations, and extrapolation techniques adds to the inherent uncertainties of oil spill advection and diffusion, which relies on a random tensor to introduce stochasticity in the movement of Lagrangian particles. The Bayesian optimization method, through its acquisition function, seeks to minimize the score without directly accounting for the specific oceanic conditions in which the oil spill simulations were conducted. Given that the currents in the affected area may be weaker than they should be, advection is likely underestimated, preventing the slick from being transported as far north as it would in real conditions. As a result, diffusion becomes the dominant process in achieving a better match between observations and simulations.  
Beyond the lack of information on slick thickness, another observational limitation stems from the use of three different satellite sensors, which could influence the results due to variations in oil slick detection across sensors. The only set of experiments using the same sensor was from observation O3 to observation O4, which also coincided with the best results in terms of the CSS metric.  
Furthermore, when comparing the observed evolution of the oil slick with FSS and CSS metrics, there are cases where the modeled slick performs well in one metric but not in the other. This was particularly evident in observations O5 and O6, where the slick area was larger than in previous observations. While the optimization improved the FSS metric for these cases, the same improvement was not reflected in the CSS metric.

\subsection{Considerations on experimental settings}\label{subsec4.2}
All experiments were conducted as sequential runs on a compute node of the JUNO Hybrid Cluster at the High-Performance Computing Center of the Euro-Mediterranean Center on Climate Change (CMCC)\footnote{https://www.cmcc.it/what-we-do/high-performance-computing-center-hpcc}.  
One of the key motivations for proposing OSO and MSO was the high computational cost of a single optimization process. In the worst-case scenario for the analyzed event, a single model run required approximately 295.54 seconds ($\sim$5 minutes) of CPU time, with a maximum memory usage of 292 MB and an average of 228.83 MB. During the training process, CPU time increased to 26,970 seconds ($\sim$7.49 hours), with a peak memory usage of 568 MB and an average of 484.36 MB. The OSO procedure involved one training step and four single model runs, whereas the more computationally demanding MSO required five training steps, further increasing resource consumption.  
For operational responses to oil spill incidents, computational time and resource usage could be limiting factors, as optimizing model performance may come at the expense of a rapid response. However, OSO demonstrated that even a single optimization step was sufficient to achieve improvements across all evaluated metrics.  
An added advantage of the proposed coupled workflow is its ability to provide an integrated solution for parameter selection, model execution, and result generation. This process can be easily automated, for instance, when a new observation becomes available, thereby reducing overall response time.  
Lastly, optimizing for FSS did not automatically lead to improvements in the CSS metric. However, the flexibility of the system allows the optimization metric to be modified, potentially yielding a different combination of MEDSLIK-II parameters better suited for specific evaluation criteria.

\subsection{Open Questions, Limitations, and Future Directions}\label{subsec4.3}
One question that remains unanswered in this study is which metric is most useful for characterizing oil slicks under different conditions. For example, would optimizing CSS better represent oil slicks in a coastal regime, as considered in this study? Conversely, how would the method perform for offshore slicks exhibiting tiger-tail features? While this remains unclear, the proposed approach allows for further exploration of these questions, even within the same set of experiments presented in this study.  
A key limitation of the system is that it does not account for the amount of oil being beached, nor does it consider oil in subsurface or sedimented compartments. It remains uncertain how the optimization process could be effectively applied to these aspects. Consequently, our approach focuses on optimizing surface oil concentrations, specifically in terms of centroid and shape, while disregarding mass balance considerations.  
Despite these limitations, the results of this study demonstrate that Bayesian optimization successfully enhanced simulation accuracy. On average, the FSS metric improved by 5.82\% with OSO and 11.07\% with MSO, while for CSS, the improvements were 5.97\% and 4.21\%, respectively. These results highlight that even a single optimization step can improve the modeled oil slick patterns in future simulations, as biases in the forcing data used for MEDSLIK-II become better incorporated into the optimization strategy.  
The findings also indicate that further experiments could introduce additional complexity to the system by modifying parameters and refining the methodology. Testing the workflow on additional real-world oil spill events will be essential to assess whether similar or even greater improvements can be achieved under different environmental conditions, ultimately demonstrating the generalizability of the approach.  
Despite these constraints, this study successfully demonstrates an automated method for improving oil spill simulations with MEDSLIK-II using Bayesian optimization, offering a systematic and efficient way to enhance model accuracy.

\section{Conclusions}\label{sec5}
In this study we propose a novel approach by using BayesOpt with the MEDSLIK-II model to simulate the oil slick behavior in Baniyas Thermal Station accident in 2021. We demonstrated that such an optimization framework can be effectively coupled with numerical simulation to increase the accuracy of the results when considering the FSS metric.
It is important to remark that, while the experimental evaluation presents a single study area using only a part of the satellite imagery available for the same accident, the framework is flexible enough for testing a variety of optimization scenarios. For example, it could be applied to different sets of environmental data (currents, winds and sea surface temperature) and considering additional MEDSLIK-II model parameters that were not tested in this first experiment.
The capacity to change and test different components of the framework opens up further enhancement possibilities for this same experiment and for other real world scenarios. Moreover, the workflow presented in this study is suitable for being used in an operational setting, in particular when considering the OSO approach.
A promising direction for future work is to develop a unified objective function that integrates multiple metrics, such as FSS and CSS, allowing the optimization process to extract more comprehensive information from multiple evaluations simultaneously. This approach would enhance the robustness of the optimization by balancing different performance aspects.
Future developments could also include relocating the framework to other oil spill events to assess its generalizability, which will ensure that the framework is reliable in an operational environment. Moreover, further evaluation of the approach by testing different acquisition functions to potentially improve the optimization process, expanding the temporal horizon to capture longer-term spill dynamics, and extending the number of parameters subject to optimization (we currently focus on only two) could also be explored. Additionally, investigating alternative parametrizations might further refine simulation accuracy.


\section*{Conflicts of Interest} 
The authors declare no conflict of interest.



\section*{Funding}
This work has been in part supported by the iMagine project. The project iMagine receives funding from the European Union’s Horizon Europe research and innovation program under grant agreement number 101058625.

\section*{Data Availability}

The data used in this study is available on Zenodo at: \url{https://zenodo.org/records/11354663}

\section*{Acknowledgments}

G. Accarino's research has been supported by the NSF through the Learning the Earth with Artificial intelligence and Physics (LEAP) Science and Technology Center (STC) (Award \#2019625). M. De Carlo has been also supported in this work by OGS and CINECA under HPC-TRES award number 2023-05.


\printbibliography


\renewcommand\theequation{\Alph{section}\arabic{equation}} 
\counterwithin*{equation}{section} 
\renewcommand\thefigure{\Alph{section}\arabic{figure}} 
\counterwithin*{figure}{section} 
\renewcommand\thetable{\Alph{section}\arabic{table}} 
\counterwithin*{table}{section} 

\newpage
\begin{appendices}
\section*{Appendix A. Supplementary Data}\label{appendixA}
\renewcommand{\appendixname}{Supplementary Material}
\renewcommand{\thefigure}{S\arabic{figure}} \setcounter{figure}{0}
\renewcommand{\thetable}{S\arabic{table}}\setcounter{table}{0}

\begin{table}[h]
\caption{The MEDSLIK-II parameters used in the simulations include the fixed parameters from the control setting, along with those optimized for the CSS and FSS metrics. In the OSO approach, optimization was conducted only between observations O1 and O2, while in the MSO approach, optimization was performed iteratively across all subsequent observations.\label{tab:tableS1}}
    \centering
    \begin{tabular}{lcc}
        \hline
        & \textbf{Horizontal Diffusivity} ($K_h$) (m$^2$/s) & \textbf{Wind Drift} ($\alpha$) (m/s) \\
        \hline
        \textbf{Control setting} & 2.00 & 0.00 \\
        \textbf{OSO O2} & 10.26 & 0.01 \\
        \textbf{MSO O3} & 8.54 & 0.01 \\
        \textbf{MSO O4} & 12.68 & 0.00 \\
        \textbf{MSO O5} & 2.33 & 0.00 \\
        \textbf{MSO O6} & 4.50 & 0.04 \\
        \hline
    \end{tabular}
\end{table}

\begin{figure}[htbp]
    \centering
    \includegraphics[width=\textwidth]{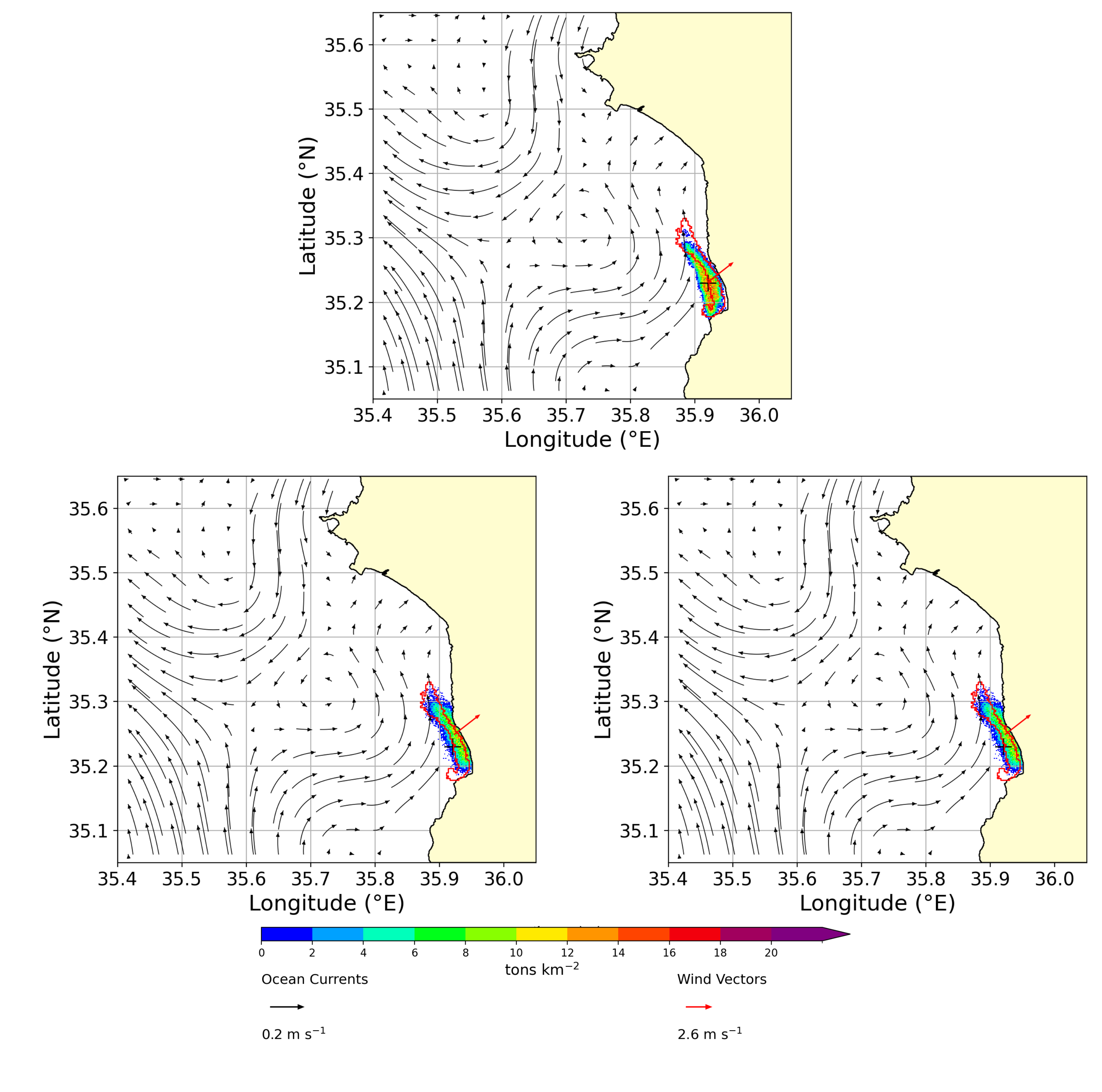}
    \caption{MEDSLIK-II surface oil concentrations starting from O1 (24/08/2021 03:43) as initial condition and stopping at O2 (24/08/2021 08:10). The upper panel represents control simulation with default physical parameters. Bottom left panel represents the optimized simulation using the OSO (left) and MSO (right) settings.}
    \label{fig:figureS1}
\end{figure}

\begin{figure}[htbp]
    \centering
    \includegraphics[width=\textwidth]{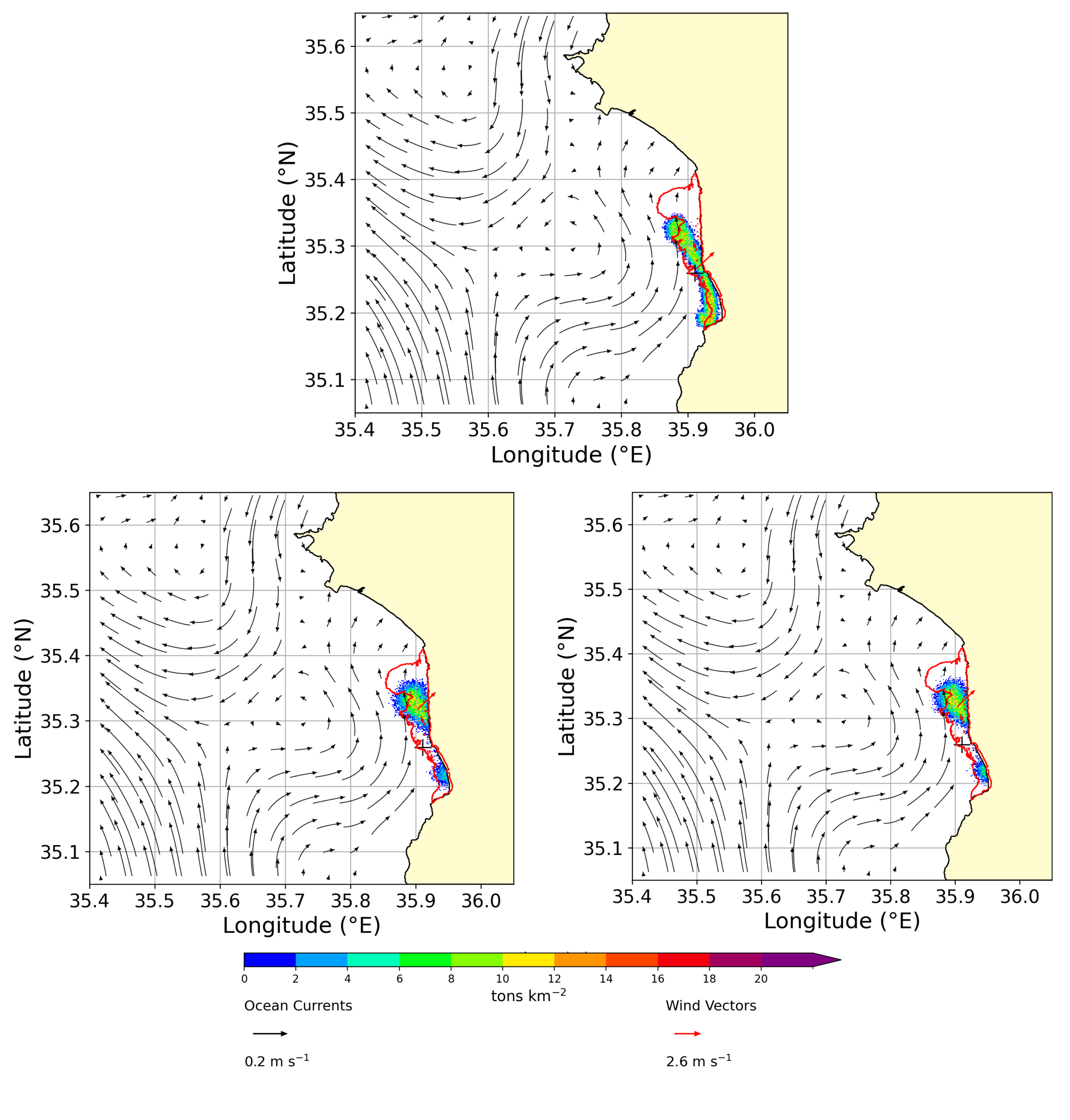}
    \caption{MEDSLIK-II surface oil concentrations starting from O2 (24/08/2021 08:10) as initial condition and stopping at O3 (24/08/2021 15:33). The upper panel represents control simulation with default physical parameters. Bottom left panel represents the optimized simulation using the OSO (left) and MSO (right) settings.}
    \label{fig:figureS2}
\end{figure}

\begin{figure}[htbp]
    \centering
    \includegraphics[width=\textwidth]{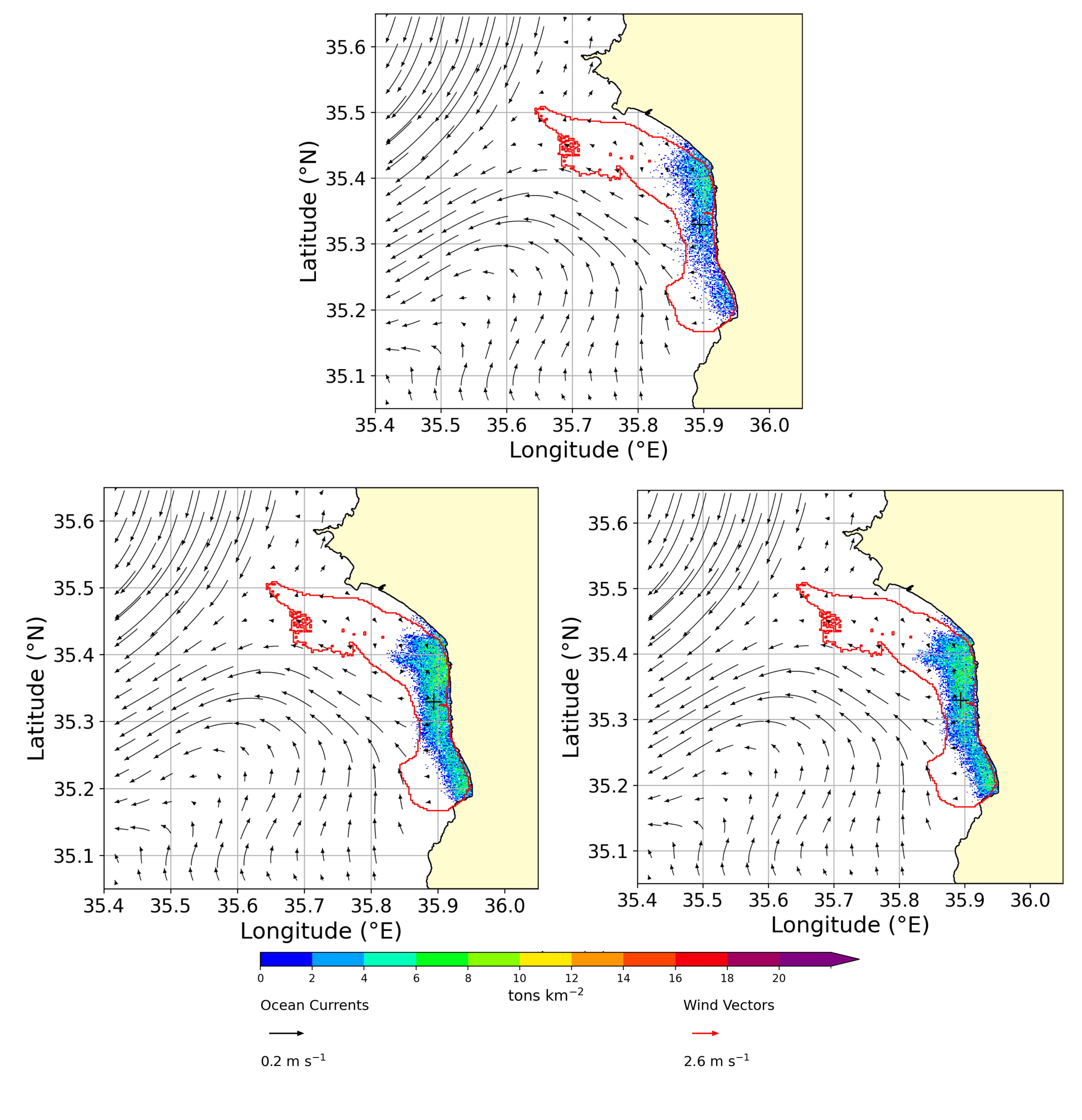}
    \caption{MEDSLIK-II surface oil concentrations starting from O4 (25/08/2021 03:34) as initial condition and stopping at O5 (26/08/2021 07:57). The upper panel represents control simulation with default physical parameters. Bottom left panel represents the optimized simulation using the OSO (left) and MSO (right) settings.}
    \label{fig:figureS3}
\end{figure}

\begin{figure}[htbp]
    \centering
    \includegraphics[width=\textwidth]{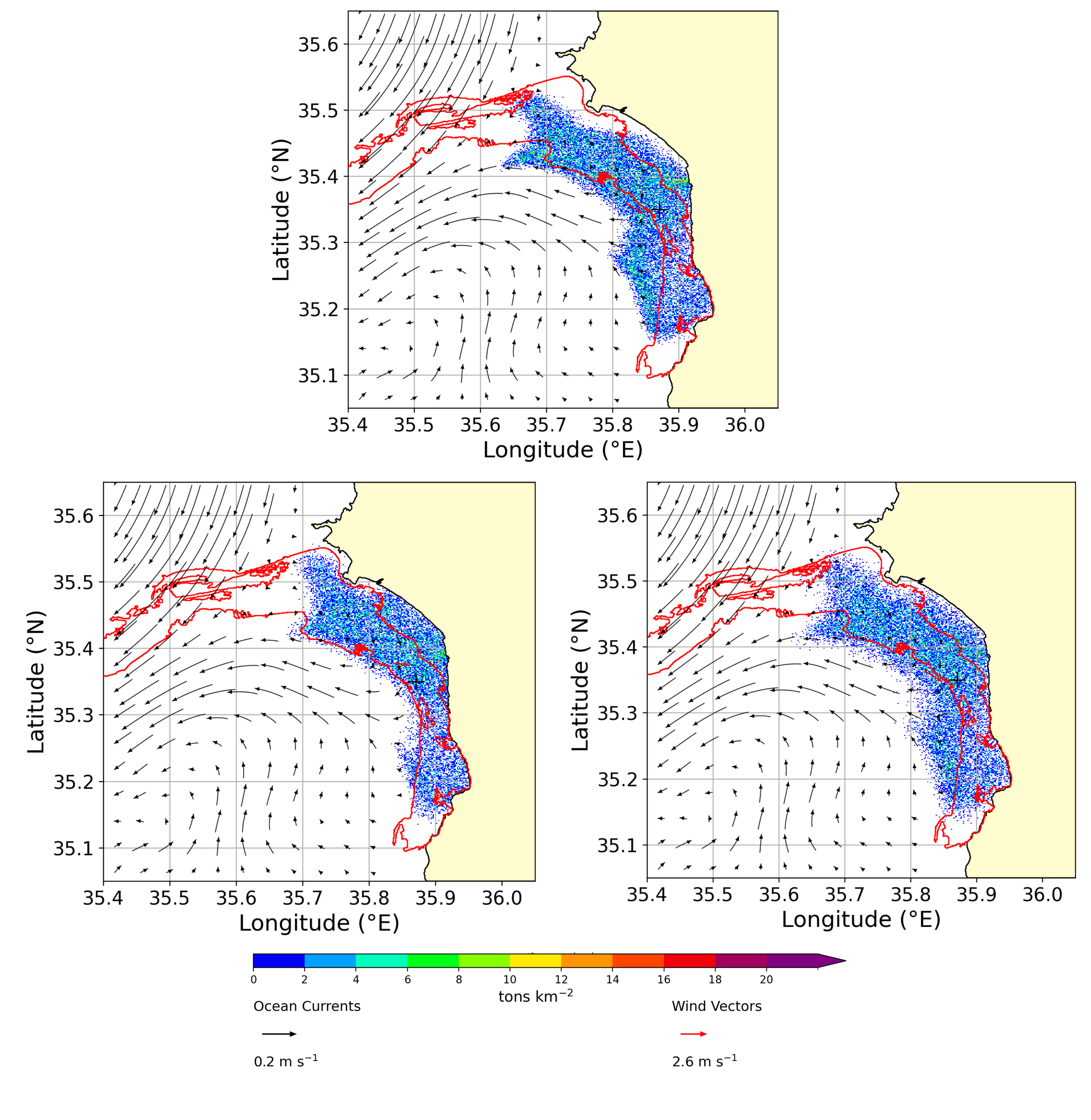}
    \caption{MEDSLIK-II surface oil concentrations starting from O5 (26/08/2021 07:57) as initial condition and stopping at O6 (27/08/2021 08:10). The upper panel represents control simulation with default physical parameters. Bottom left panel represents the optimized simulation using the OSO (left) and MSO (right) settings.}
    \label{fig:figureS4}
\end{figure}

\end{appendices}

\end{document}